\documentclass{elsarticle}
\usepackage[utf8]{inputenc}
\usepackage[T1]{fontenc}
\usepackage{float}
\usepackage{subfig}
\usepackage{graphicx}
\usepackage{hyperref}
\usepackage{csvsimple}
\usepackage{multirow}
\usepackage[left]{lineno} 
\usepackage{amsmath}
\usepackage{booktabs}
\usepackage{rotating}
\usepackage{tabularx}
\usepackage{booktabs}

\title{Detecting Network Instability via Multiscale Detrended Cross-Correlations and MST Topology}
\author[1]{Jose De Leon Miranda}
\author[1,2]{Marina Dolfin\corref{cor1}}
\author[1]{George Kapetanios}
\author[1]{Leone Leonida}

\cortext[cor1]{Corresponding author}
\ead{marina.dolfin@kcl.ac.uk}

\address[1]{King's College London, King's Business School, London, WC2B 4BG, UK}
\address[2]{University of Messina, Department of Engineering, Messina, 98166, IT}

\begin{document}
\begin{keyword}
Multiscale networks \sep Detrended Cross-Correlation Analysis (DCCA) \sep Minimum spanning trees \sep correlation networks \sep scaling \sep network instability
\end{keyword}

\begin{abstract}
We introduce a multiscale measure of network instability based on the joint use of
Detrended Cross-Correlation Analysis (DCCA) and Minimum Spanning Tree (MST)
filtering. The proposed metric—the \emph{Elastic Detrended Cross-Correlation Ratio} (Elastic DCCR)—is defined as a finite-difference measure of the logarithmic sensitivity of the average MST length to the observation scale. It captures how the structure of cross-correlation networks deforms across different investment horizons.
When applied to a network of global equity indices, the Elastic DCCR rises sharply during
episodes of financial stress, reflecting increased short-term coordination among
investors and a contraction of correlation distances. The measure reveals scale-dependent reconfigurations in network topology that
are not visible in single-scale analyses, and highlights clear differences
between stressed and stable market regimes. The approach does not assume covariance stationarity and relies only on scale-dependent detrended correlations; as a result, it is broadly applicable to other complex systems in which interaction strength varies with scale.

\end{abstract}

\maketitle

\section*{Introduction}
Key financial activities—including portfolio construction, risk management, and
the pricing of financial instruments—critically depend on the analysis of
connectedness among assets and markets
\cite{Diebold2009, Barunik2018, Cont2001, Kritzman2003, Serletis2002,
Markowitz1952, Bauwens2009, Engle2012}.  
Understanding the structure and temporal evolution of such connectedness is
central to identifying regularities and stylized facts in market dynamics.
Empirical evidence shows that cross-market correlations fluctuate markedly over
time as a result of technological and financial innovations, structural breaks,
and changes in global conditions.  
These fluctuations become especially pronounced during episodes of elevated
volatility, when markets display enhanced synchronization
\cite{Podobnik:2012, Sornette}.  
In such regimes, effective dimensionality contracts, liquidity preferences
shift, and market participants increasingly coordinate their behaviour across
similar horizons \cite{Kristoufek2012, Kristoufek2013, Peters1994}.  
The resulting increase in cross-market coupling amplifies the channels through
which shocks propagate, thereby heightening systemic risk
\cite{Adrian2016, Diebold2012, Forbes2002, Kritzman2003, Bauwens2009}.

Characterising how the structure of financial networks reorganises across time
scales, and how this affects the transmission of risk, remains a central
challenge in the study of complex systems.  
Existing approaches—such as single-scale DCCA estimators or
variance-decomposition techniques of Diebold and Yilmaz—capture important
features of dependence but do not quantify how the topology of the correlation
network itself deforms as the observation scale changes.

The principal contribution of this research is to introduce a scale-comparison
metric that detects structural transitions in correlation networks by examining
how their topology changes across time horizons.  
We combine Detrended Cross-Correlation Analysis (DCCA), which measures
scale-dependent long-range correlations
\cite{Podobnik:2008, Zhou2003, Bunde2005, Yang2012, Wang2015}, with the Minimum
Spanning Tree (MST), which filters noisy interactions while retaining the
essential backbone of the network.  
We define a new quantity—the Elastic Detrended Cross-Correlation Ratio (Elastic DCCR)—as the finite-difference approximation of the derivative of the logarithmic MST average length with respect to the logarithm of the observation scale. 
This ratio provides a compact, dimensionless indicator of multiscale
topological deformation and serves as a diagnostic of instability: when the
network becomes more synchronised at short horizons than at long horizons, the
Elastic DCCR exhibits marked deviations from its scale-invariant baseline behaviour.

Our analysis shows that the Elastic DCCR captures periods of structural
transition in global equity markets, where investors compress their
interactions onto short time scales during episodes of stress.  
The metric uncovers changes in connectedness that remain hidden to
single-horizon analyses, and its behaviour contrasts systematically with the
connectedness measures of Diebold and Yilmaz \cite{Diebold2014}.  
By combining multiscale cross-correlations with topological filtering, this
framework provides a physically interpretable and computationally efficient
tool for monitoring systemic change in complex networks. In this sense, the Elastic DCCR plays the role of an order-parameter-like indicator for scale-induced topological transitions in correlation networks.

\section{Methods}
\label{Methods}
\subsection{Multiscale detrended cross-correlations for network construction}

We analyze daily financial market indices and transform the price series $P_i(t)$, 
with $i=1,\dots,N$ markets and $t=1,\dots,T$ observations, into logarithmic 
returns defined as
\begin{equation}
r_i(t) = \ln P_i(t) - \ln P_i(t-1).
\label{eq:return}
\end{equation}

To account for volatility clustering and remove heteroskedasticity at the level of each single time series, we filter the 
returns using a GARCH(1,1) model,
\begin{equation}
r_i(t) = \sigma_i(t)\,\varepsilon_i(t),
\end{equation}
and work with the standardized residuals
\begin{equation}
r_{i,f}(t) = \frac{r_i(t)}{\sqrt{h_i(t)}},
\label{eq:garch}
\end{equation}
where $h_i(t)$ is the conditional variance obtained from the GARCH fit.  
This step ensures that scale-dependent statistics reflect intrinsic temporal 
structure rather than volatility bursts.

The filtered series are then normalized to zero mean and unit variance,
\begin{equation}
\tilde{r}_i(t) = \frac{r_{i,f}(t) - \langle r_{i,f} \rangle}
{\sigma_{r_i}},
\label{eq:standardized}
\end{equation}
where $\sigma_{r_i} = \sqrt{ \langle r_{i,f}^2 \rangle - \langle r_{i,f} \rangle^2 }$ 
and $\langle \cdot \rangle$ denotes a time average over the sample period.

The above standardized returns $\tilde{r}_i(t)$ constitute the inputs for the 
Detrended Cross-Correlation Analysis (DCCA). The DCCA coefficient quantifies scale-dependent correlation between two 
non-stationary time series $x$ and $y$.  
Given a scale $s$, the coefficient is defined as
\begin{equation}
\rho_{\mathrm{DCCA}}(s) = 
\frac{F^2_{\mathrm{DCCA}}(s)}
{F_{\mathrm{DFA},x}(s)\,F_{\mathrm{DFA},y}(s)},
\label{eq:dcca}
\end{equation}
where $F_{\mathrm{DFA},x}(s)$ and $F_{\mathrm{DFA},y}(s)$ are DFA fluctuation 
functions and $F^2_{\mathrm{DCCA}}(s)$ is the detrended covariance function.

Following the DFA/DCCA framework introduced for a single time series by Peng et al.~\cite{Peng1994}
and generalized by Podobnik and Stanley~\cite{Podobnik:2008},
each time series is integrated to form its cumulative profile,
\begin{equation}
X(t) = \sum_{k=1}^{t} (x_k - \langle x \rangle),
\qquad
Y(t) = \sum_{k=1}^{t} (y_k - \langle y \rangle),
\end{equation}
and partitioned into $N_s = \lfloor T/s \rfloor$ non-overlapping segments of length $s$.
Within each segment of length $s$, a local polynomial trend is subtracted to obtain the
detrended variances
\begin{equation}
f_{\mathrm{DFA},x}^2(n,s) =
\frac{1}{s}\sum_{t=1}^{s} [X(t) - \hat{X}(t)]^2,
\qquad
f_{\mathrm{DFA},y}^2(n,s) =
\frac{1}{s}\sum_{t=1}^{s} [Y(t) - \hat{Y}(t)]^2,
\end{equation}
and the detrended covariance
\begin{equation}
f_{\mathrm{DCCA}}^2(n,s) =
\frac{1}{s}\sum_{t=1}^{s}
[X(t) - \hat{X}(t)][Y(t) - \hat{Y}(t)].
\label{eq:cov}
\end{equation}
where $\hat{X}(t)$ and $\hat{Y}(t)$ denote the fitted local polynomial trends.
Averaging over all segments yields the multiscale fluctuation and covariance functions,
\begin{equation}
F_{\mathrm{DFA},x}(s) =
\sqrt{\frac{1}{N_s}\sum_{n=1}^{N_s} f_{\mathrm{DFA},x}^2(n,s)},
\qquad
F^2_{\mathrm{DCCA}}(s) =
\frac{1}{N_s}\sum_{n=1}^{N_s} f_{\mathrm{DCCA}}^2(n,s),
\end{equation}
which together define $\rho_{\mathrm{DCCA}}(s)$ in Eq.~\eqref{eq:dcca}.

The DCCA method is well suited for detecting multiscale dependencies, long-range 
memory, and non-stationary structure in financial and other complex systems 
\cite{Podobnik:2008, Zhou2003, Kristoufek2013, Wang2015}.

\subsection{Average MST length as a multiscale connectivity measure}

The DCCA coefficient $\rho_{\mathrm{DCCA}}^{ij}(s)$ is mapped into a correlation-based distance
following the geometric construction of Mantegna~\cite{Mantegna1999},
\begin{equation}
d_{\mathrm{DCCA}}^{ij}(s) =
\sqrt{\,2 \left[1 - \left(\rho_{\mathrm{DCCA}}^{ij}(s)\right) \right]},
\label{eq:dcca_distance}
\end{equation}
which yields a symmetric, non-negative distance matrix suitable for hierarchical
clustering and minimum spanning tree filtering.
The transformation is well defined when $|\rho_{\mathrm{DCCA}}^{ij}(s)| \le 1$;
however, strong heterogeneity in long-range dependence (e.g., markedly different
Hurst exponents across series) may lead to violations of this condition.
In such cases, generalized normalization procedures may be required to ensure
distance compatibility for network construction.
 (see Research Perspectives).

We construct a time- and scale-dependent weighted network with $N$ market indices
as nodes and DCCA distances as link weights.
For each pair $(i,j)$, the distance at scale $s$ and time $t$ is
\begin{equation}
d^{ij}_{\mathrm{DCCA}}(s,t)
= \sqrt{\,2 \left[1 - \left(\rho^{ij}_{\mathrm{DCCA}}(s,t)\right)\right]}.
\end{equation}

The resulting weighted, fully connected network is filtered using the
Minimum Spanning Tree (MST) filtering procedure
\cite{Mantegna1999, Onnela2003, Tumminello2005, Bonanno2004}.
 
The MST preserves the strongest connections while eliminating cycles, yielding a sparse backbone that captures the essential topological structure of the underlying correlation network.
In our analysis, the MST is constructed using Kruskal’s algorithm~\cite{Kruskal1956}.

To quantify the degree of connectivity in the system at scale $s$ and time $t$, 
we compute the \emph{average tree length},
\begin{equation}
L(s,t) = \frac{1}{N-1}
\sum_{(i,j)\in \mathrm{MST}^t}
d^{ij}_{\mathrm{DCCA}}(s,t),
\label{eq:tree_length}
\end{equation}
where the sum runs over the $N-1$ edges of the MST.  
Shorter values of $L(s,t)$ indicate stronger clustering and tighter interdependence among markets, whereas larger values reflect more weakly connected or more dispersed network structures.~\cite{Mantegna1999, Tumminello2005}.  
Thus, $L(s,t)$ provides a compact descriptor of the network’s multiscale 
connectivity.

\subsection{Scaling structure of the MST topology and Elastic DCCR}

Because Detrended Cross-Correlation Analysis yields scale-dependent measures of dependence, the geometry of the corresponding correlation network is also expected to vary with the observation scale. In systems characterised by heterogeneous interaction horizons, such as financial markets with coexisting short- and long-term investors, it is natural to ask whether this scale dependence follows a regular scaling structure.

A convenient summary of network geometry at scale $s$ and time $t$ is provided by the average length of the Minimum Spanning Tree, $L(s,t)$. If correlations reorganise smoothly across horizons, the variation of $L(s,t)$ with respect to $s$ may be approximated locally by a power-law relation,
\begin{equation}
L(s,t) \propto s^{\alpha(t)},
\label{eq:powerlaw_L}
\end{equation}
where $\alpha(t)$ is a time-varying scaling exponent. This relation represents a form of local scale invariance in the network topology: small values of $\alpha(t)$ indicate weak sensitivity of connectivity to scale, while larger or rapidly changing values reflect pronounced multiscale heterogeneity. From a network perspective, such behaviour suggests that correlation distances may reorganise in an approximately self-similar manner across observation scales, motivating the exploration of scaling relations for the average MST length.

Under the scaling hypothesis in Eq.~\eqref{eq:powerlaw_L}, the logarithm of the MST length is approximately linear in $\log s$,
\begin{equation}
\log L(s,t) = \alpha(t)\log s + C(t),
\label{eq:log_linear}
\end{equation}
where $C(t)$ is a slowly varying intercept. Deviations from linearity in log-log space correspond to departures from scale-invariant organisation and signal distortions in the multiscale structure of the correlation network.

Direct estimation of $\alpha(t)$ requires fitting across multiple scales and is therefore sensitive to the choice of scale range, window length, and noise. For monitoring purposes, a simpler and more robust approach is to estimate the local logarithmic sensitivity of $L(s,t)$ using a finite-difference approximation across two representative horizons.

We therefore define the \emph{Elastic Detrended Cross-Correlation Ratio} (Elastic DCCR) as
\begin{equation}
\text{Elastic DCCR}(t)
=
\frac{\log L(s_{\text{long}},t) - \log L(s_{\text{short}},t)}
{\log s_{\text{long}} - \log s_{\text{short}}},
\label{eq:elastic_dccr}
\end{equation}
where $s_{\text{short}} < s_{\text{long}}$. This quantity provides a dimensionless, scale-invariant measure of how network connectivity deforms between short and long observation horizons.

When the power-law approximation in Eq.~\eqref{eq:powerlaw_L} holds locally, the Elastic DCCR coincides with the scaling exponent,
\begin{equation}
\text{Elastic DCCR}(t) \approx \alpha(t),
\label{eq:dccr_alpha}
\end{equation}
and thus serves as a finite-difference estimate of $\partial \log L / \partial \log s$. Importantly, this interpretation does not require global scale invariance: the Elastic DCCR remains well defined even when the scaling relation holds only approximately or over limited scale ranges.

Sharp temporal variations in the Elastic DCCR indicate local breakdowns of scale invariance and reflect abrupt reconfigurations of network topology. Such events are particularly relevant during periods of market stress, when short-horizon correlations tend to increase rapidly due to liquidity shocks or shifts in risk sentiment, while longer-horizon dependencies adjust more gradually. This scale asymmetry produces transient distortions in the MST geometry, which manifest as spikes or sign reversals in the Elastic DCCR.

By focusing on relative changes across scales rather than absolute levels of connectivity, the Elastic DCCR provides a compact and interpretable diagnostic of multiscale structural instability in correlation networks.

\section{Monitoring multiscale network dynamics in financial markets}

\subsection{Data}

We analyze daily closing prices of major equity indices for the G7 countries, China, and Russia. 
The dataset includes the S\&P~500 (United States), S\&P/TSX Composite (Canada),
CAC~40 (France), DAX (Germany), FTSE~MIB (Italy), Nikkei~225 (Japan),
FTSE~100 (United Kingdom), Hang Seng Index (China), and the MOEX Russia Index. 
The sample spans the period from 5 March 2013 to 30 May 2023.  
All data and code---including automated retrieval from Yahoo Finance---are 
available at  
\href{https://github.com/JoseDLM/Investor-Behavior-and-Multiscale-Cross-Correlations}
{https://github.com/JoseDLM/Investor-Behavior-and-Multiscale-Cross-Correlations}.
Daily prices are converted into logarithmic returns and filtered for
heteroskedasticity using a GARCH(1,1) model.
Detrended Cross-Correlation Analysis (DCCA) is then applied to the filtered and
standardized return series across a range of scales, following the methodology
outlined in Section~\ref{Methods}.
Scale-dependent DCCA coefficients are then transformed into distances using a
nonlinear mapping inspired by the construction proposed by Mantegna\cite{Mantegna1999}, enabling a geometric representation suitable
for Minimum Spanning Tree (MST) construction.

To illustrate the multiscale evolution of cross-market dependencies, we examine
the DCCA distances between each index and the S\&P~500, taken as a benchmark,
over time and across scales.

\begin{figure}[H]
\centering
\includegraphics[width=\textwidth]{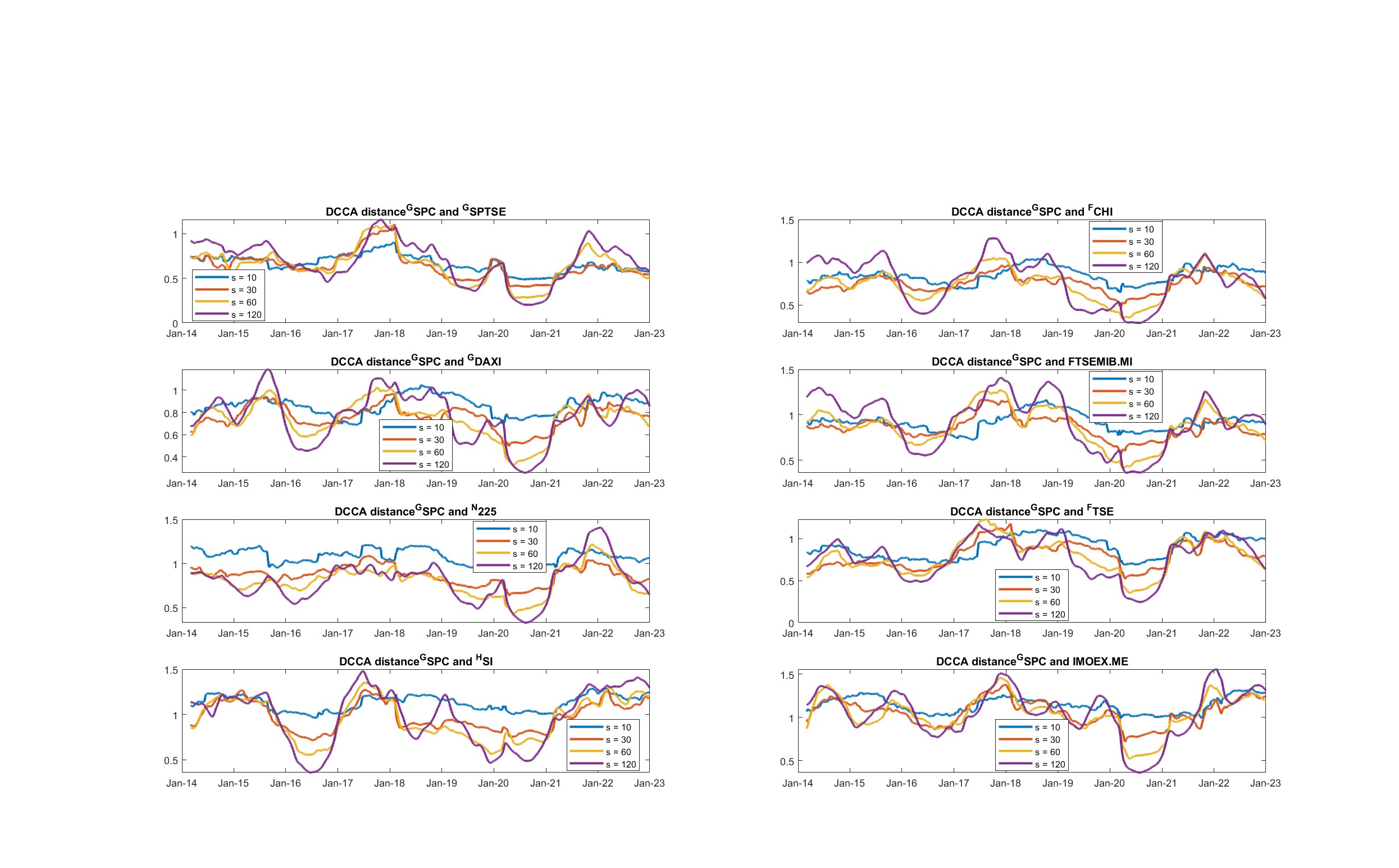}
\caption{DCCA distances between the S\&P~500 and other indices, computed across
DCCA scales ranging from $s = 10$ to $120$ days using a rolling window of
$w = 250$ trading days.}
\label{fig:paircorrelations}
\end{figure}

The DCCA distances exhibit clear temporal variation, increasing during periods of
market stress and decreasing during stable periods.
Short-horizon distances respond more sharply to local fluctuations, reflecting
the rapid evolution of short-term dependencies.
Longer horizons evolve more slowly and capture persistent structural components
of cross-market interactions.

Following the onset of COVID-19, long-horizon distances increase substantially,
consistent with scale-dependent reorganization of market structure during the
transition from crisis to recovery.
Short-horizon distances remain comparatively low throughout the early recovery
phase, indicating that short-term co-movements stay elevated even as
longer-horizon structure gradually decorrelates.

To track time-varying connectedness, we compute the DCCA distance matrices
$d^{ij}_{\mathrm{DCCA}}(s,t)$ using a sliding window of width $w$ with
one-day steps.
For each $t = 1,\dots, T-w+1$, distances are estimated from observations in
the interval $[t, t+w-1]$.
These distances are then filtered through the MST to obtain the backbone of
cross-market interdependencies.

\begin{figure}[ht!]
\centering
\includegraphics[width=5.5cm]{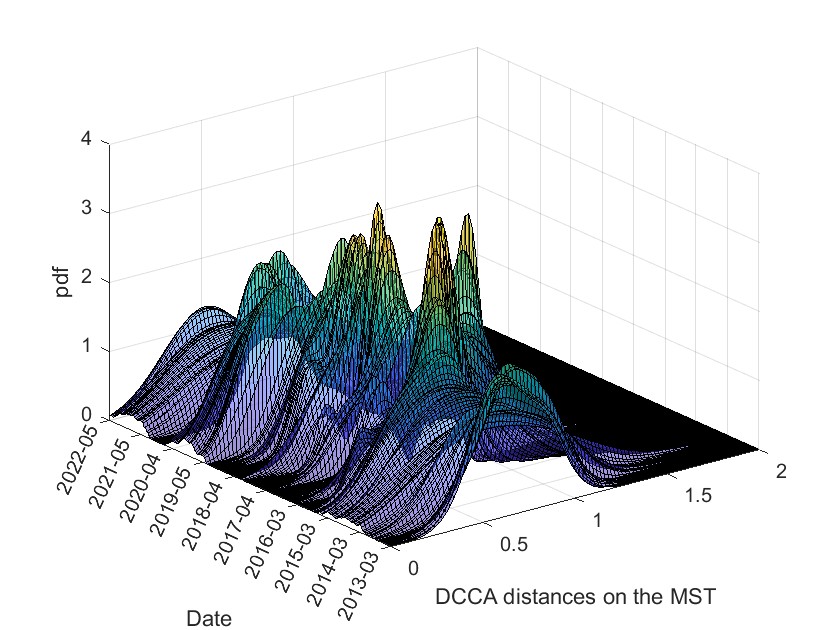}
\quad
\includegraphics[width=5.5cm]{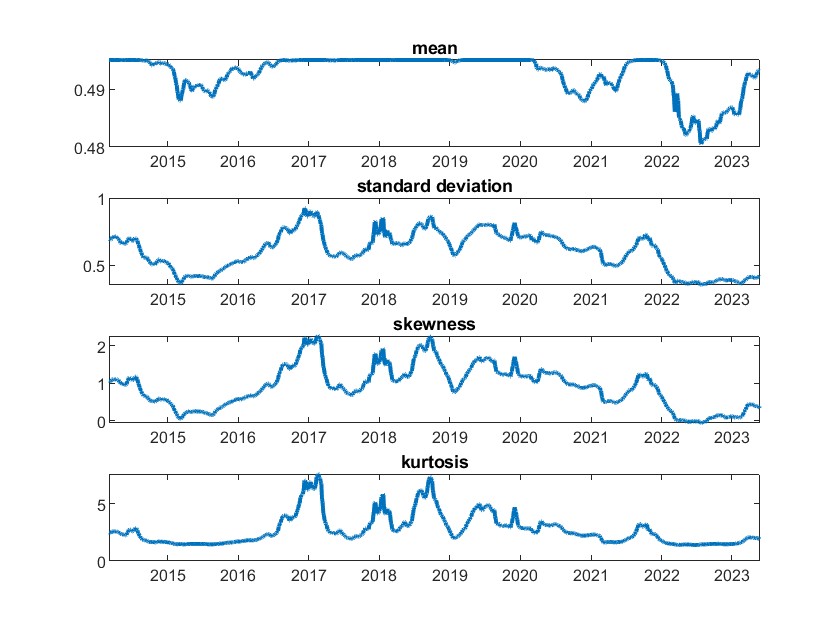}
\caption{Time evolution of the density of MST-filtered 1-month DCCA distances
(left) and the corresponding dynamics of the first four moments (right).}

\label{fig:MST30}
\end{figure}

\begin{figure}[ht!]
\centering
\includegraphics[width=5.5cm]{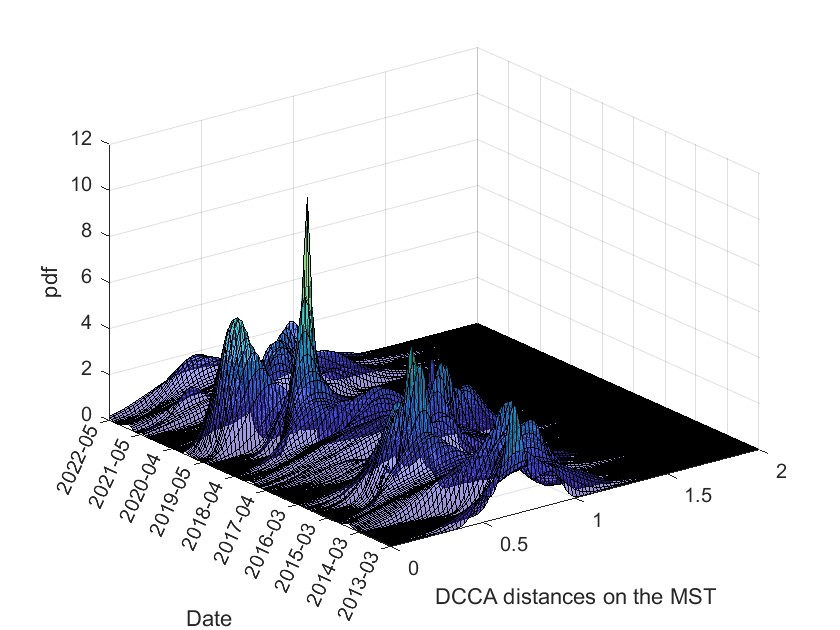}
\quad
\includegraphics[width=5.5cm]{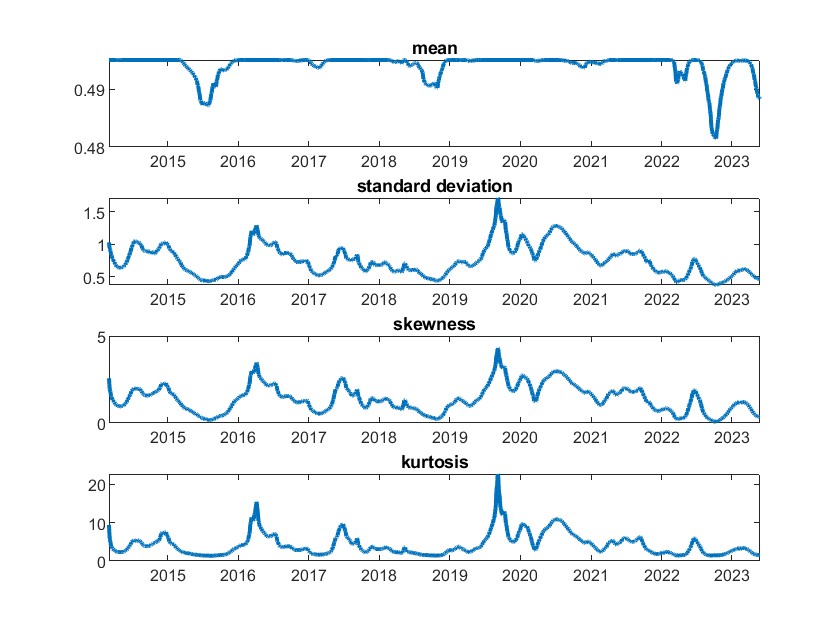}
\caption{Time evolution of the density of MST-filtered 4-month DCCA distances
(left) and the corresponding dynamics of the first four moments (right).}

\label{fig:MST120}
\end{figure}

Figures~\ref{fig:MST30} and \ref{fig:MST120} display kernel density estimates of 
the MST-filtered DCCA distances at the 1-month and 4-month horizons, respectively.
The panels on the right show the first four moments computed over the same
rolling window.

These distributional diagnostics complement the Elastic DCCR by revealing higher-order structural changes in the MST backbone that are not captured by average connectivity alone.

At shorter scales, the kurtosis of the MST-filtered distance distribution remains 
finite, although it increases markedly during turbulent periods.  
This indicates the presence of heavy—but not divergent—tails associated with 
rapid shifts in short-term cross-market structure.  
At longer scales, the distributions become more sharply peaked, with occasional 
episodes of elevated kurtosis reflecting abrupt structural adjustments.  
In all cases, the fourth moment remains empirically finite, confirming that the 
distance distributions do not exhibit power-law tails with exponent below~4.  
The reduced prevalence of extremes at longer scales is consistent with persistent 
long-range correlation structure.

Periods of instability correspond to elevated uncertainty and rapid adjustments 
in global risk conditions, producing strong short-term co-movements and 
scale-dependent distortions of network topology.  
The multiscale MST representation thus provides a compact and informative view 
of the temporal evolution of cross-market connectedness.

\section{Real versus synthetic networks: a multiscale scaling test}

To assess whether the structural changes observed in the correlation-based
networks arise from genuine cross-market dependencies rather than from
univariate volatility dynamics, we construct a synthetic benchmark based on
independent GARCH(1,1) processes.
These series reproduce volatility clustering and heteroskedasticity but contain
no cross-correlations by construction.
They therefore serve as a baseline for distinguishing multivariate dependence
from purely univariate effects.
This comparison provides a direct empirical counterpart to the scaling hypothesis introduced in Section~1.3, allowing us to assess when the average MST length follows an approximately local power-law structure and when this structure breaks down.

Our aim is to compare the scale dependence of the average MST length $L(s,t)$ in
real and synthetic data, and to identify deviations from power-law behaviour
that signal nontrivial multiscale structure.
Both systems are analysed using a sliding window of width $w=250$ trading days,
with $t = 1,\dots,T-w+1$.

\vspace{.2cm}
\noindent\textbf{Synthetic data.}
We generate $N=9$ independent GARCH(1,1) series,
\[
r_t = \sigma_t \varepsilon_t,\qquad
\sigma_t^2 = \omega + \alpha r_{t-1}^2 + \beta \sigma_{t-1}^2,
\]
with parameters $\omega = 10^{-6}$, $\alpha = 0.1$, and $\beta = 0.85$, chosen
to reflect typical financial-market persistence ($\alpha+\beta=0.95$) and
a reasonable unconditional variance.
Each series is initialized with a different random seed, and no
cross-correlations are imposed by construction.

\noindent\textbf{Real data.}
The real dataset consists of daily closing prices for the G7+China+Russia equity
indices over 2013--2023, filtered via GARCH(1,1) to remove asset-level
heteroskedasticity.
The objective is to isolate structural instabilities arising from the
cross-correlation network rather than from univariate volatility effects.

We visualize the surface $\log L(s,t)$ over the $(\log s, t)$ plane for scales
ranging from 5 to 160 days, logarithmically spaced, to highlight the joint
time–scale evolution of network connectivity.

\begin{figure}[ht]
\centering
\begin{minipage}{0.48\linewidth}
  \centering
  \includegraphics[width=\linewidth,height=5cm,keepaspectratio]{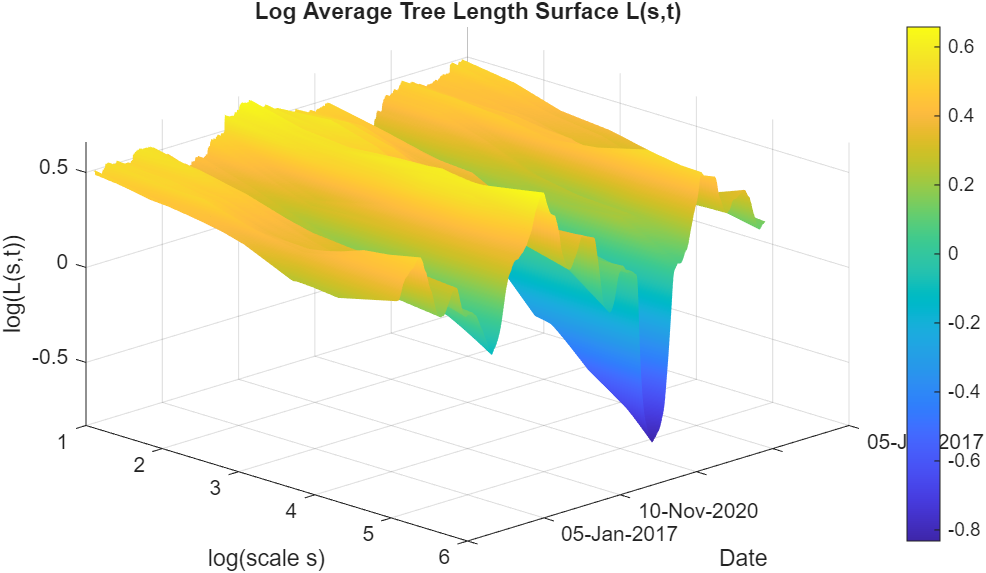}
  \caption*{(a) Real data}
\end{minipage}\hfill
\begin{minipage}{0.48\linewidth}
  \centering
  \includegraphics[width=\linewidth,height=5cm,keepaspectratio]{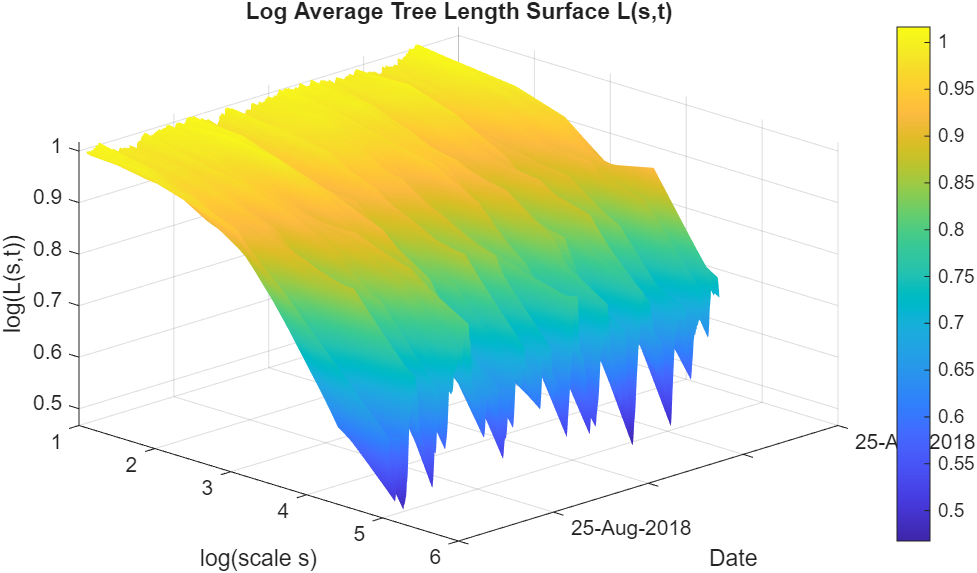}
  \caption*{(b) Synthetic GARCH data}
\end{minipage}
\caption{Surfaces of $\log L(s,t)$ over the $(\log s,t)$ plane for real (a) and
synthetic (b) data.}
\end{figure}
Real markets show frequent distortions in the $(\log s,t)$ surface, whereas the
synthetic system exhibits a smoother, nearly linear log--log structure.
This indicates that the synthetic data are consistent with an approximately
local power-law scaling, while real data exhibit scale-dependent disruptions.

The Elastic DCCR is computed for three distinct short- and long-horizon scale
pairs, as indicated in the figure legend.
\begin{figure}[ht]
\centering
\begin{minipage}{0.48\linewidth}
  \centering
\includegraphics[width=\linewidth,height=4cm,keepaspectratio]{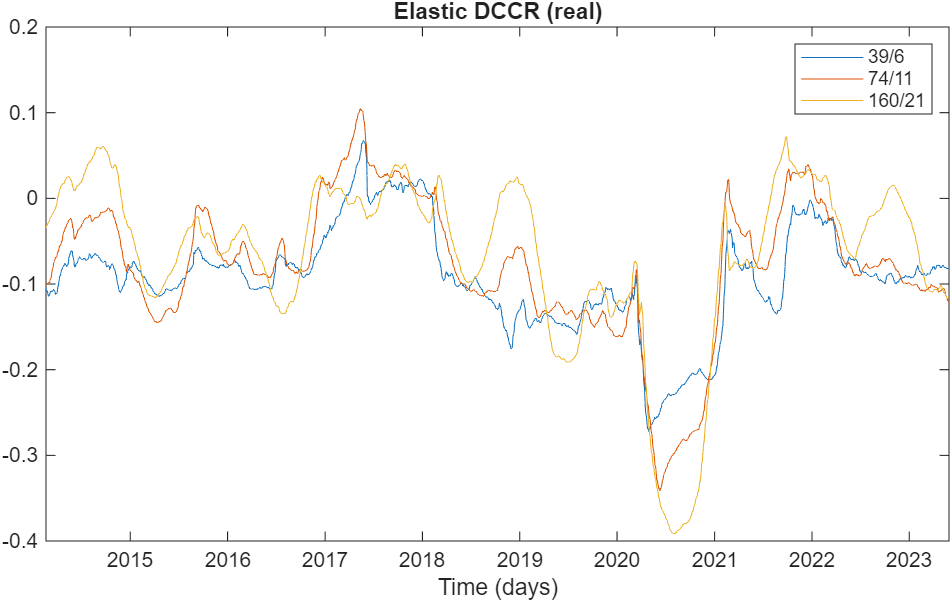}
  \caption*{(a) Real data}
\end{minipage}\hfill
\begin{minipage}{0.48\linewidth}
  \centering
 \includegraphics[width=\linewidth,height=4cm,keepaspectratio]{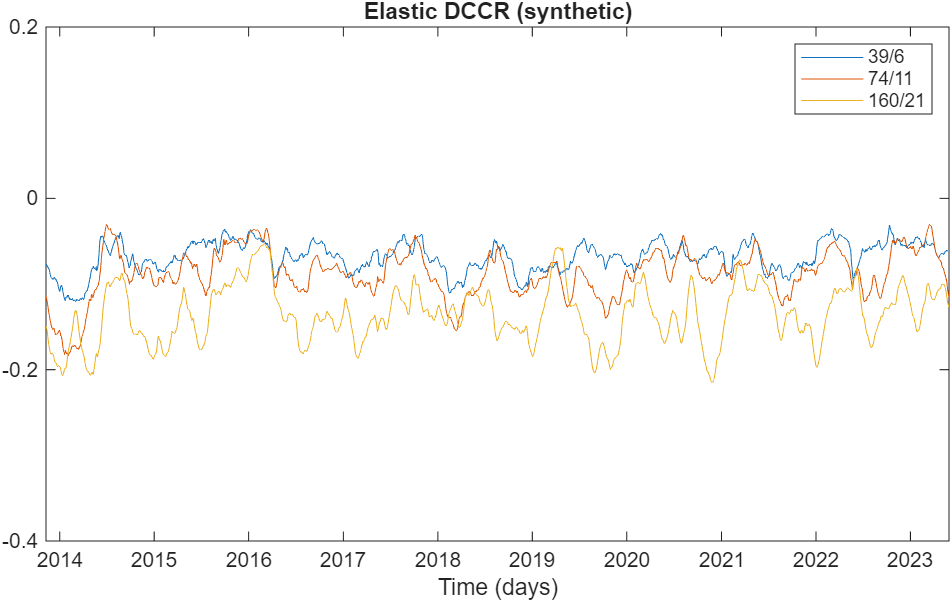}
  \caption*{(b) Synthetic data}
\end{minipage}

\end{figure}

Synthetic data show stable, nearly scale-invariant behavior, while real systems 
feature intermittent bursts, reflecting multiscale structural shifts in 
correlation networks.  
These deviations are consistent with the emergence of collective market dynamics
that extend beyond volatility persistence alone.

To further interpret the scale-dependent behavior captured by the Elastic DCCR,
we consider additional diagnostic quantities that probe how the average MST
length varies with scale.
Specifically, examining first- and second-order variations of $\log L(s,t)$ with
respect to $\log s$ provides insight into both the local elasticity of network
connectivity and the rate at which this elasticity itself changes across
horizons.
These diagnostics are introduced to complement, rather than replace, the Elastic
DCCR.

\vspace{.2cm}
\noindent\textbf{Local scaling exponent diagnostics.}
 Before examining the local scaling exponent, it is important to note that the
interpretation of $\alpha(t)$ differs fundamentally between real and synthetic
systems.
While a local exponent can be estimated in both cases, in the synthetic GARCH
benchmark it reflects a trivial, noise-induced scaling associated with
independent series.
In contrast, for real markets $\alpha(t)$ provides a diagnostic of nontrivial,
time-varying distortions in the multiscale structure of the correlation network.

We estimate the time-varying scaling exponent $\alpha(t)$ via OLS regressions of
$\log L(s,t)$ on $\log s$ within moving windows.
The resulting goodness of fit, measured by the coefficient of determination
$R^2$, varies substantially for real data but remains consistently high for the
synthetic system:

\begin{figure}[ht]
\centering
\begin{minipage}{0.48\linewidth}
  \centering
  \includegraphics[width=\linewidth,height=4.5cm,keepaspectratio]{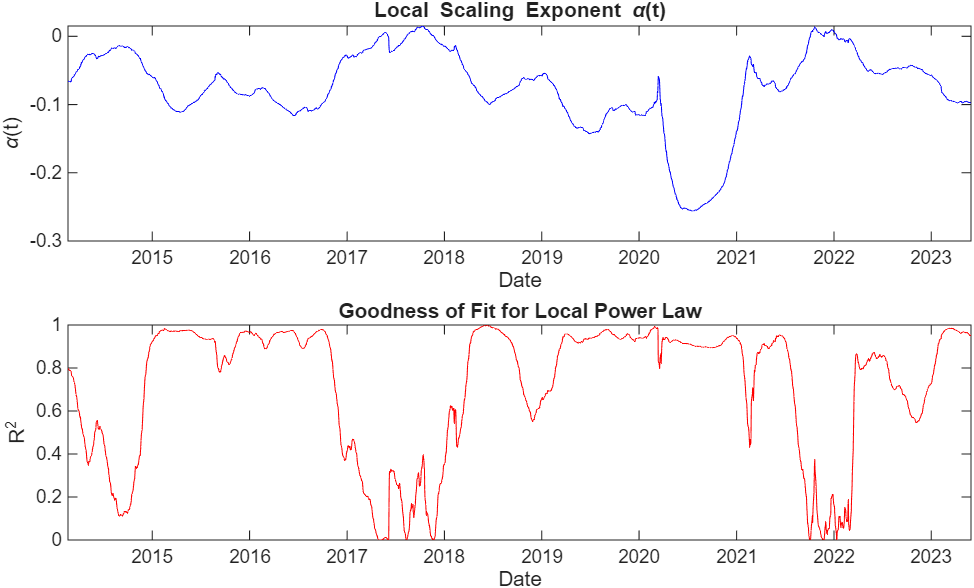}
  \caption*{(a) Real data}
\end{minipage}\hfill
\begin{minipage}{0.48\linewidth}
  \centering
\includegraphics[width=\linewidth,height=4.5cm,keepaspectratio]{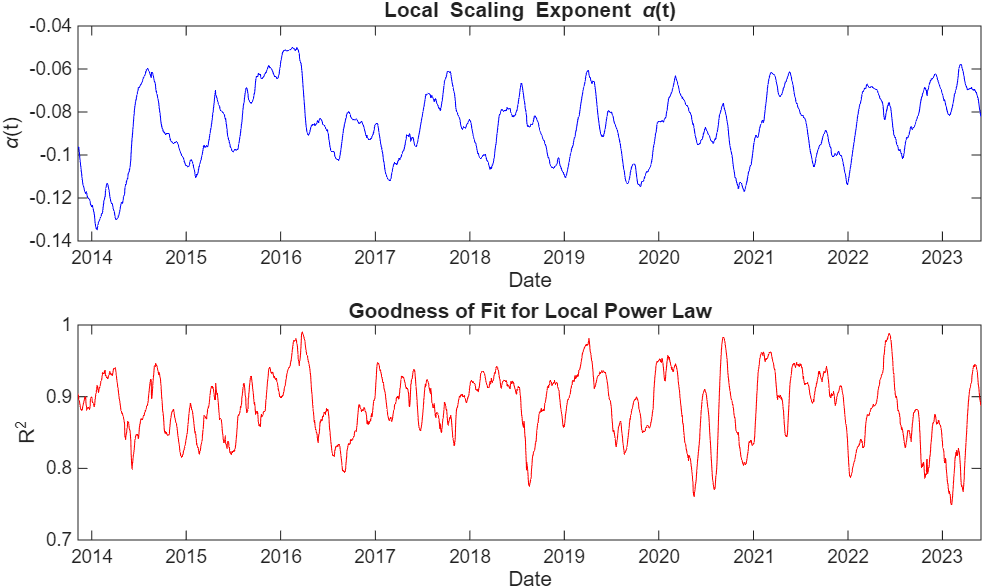}
  \caption*{(b) Synthetic data}
\end{minipage}
\caption{Estimated local scaling exponent $\alpha(t)$ and corresponding $R^2$
for real (a) and synthetic (b) data.}
\end{figure}

These results show that real markets frequently depart from linear 
scale-invariant behavior, while synthetic GARCH series maintain stable power-law 
structure.

\vspace{.2cm}
\noindent\textbf{Second-order scaling.}
To complement the first-order diagnostic based on the local slope, we also
examine curvature in the $\log L$--$\log s$ relation via a local second-order
expansion in log--log space,
\begin{equation}
\log L(s,t) = \gamma(t) + \alpha(t)\log s + \beta(t)(\log s)^2.
\label{eq:second_order_scaling}
\end{equation}

A nonzero $\beta(t)$ signals scale-dependent variations in the effective scaling
behavior and departures from simple scale invariance.

\begin{figure}[ht]
\centering
\includegraphics[
  width=0.48\textwidth,
  height=13cm,
  keepaspectratio
]{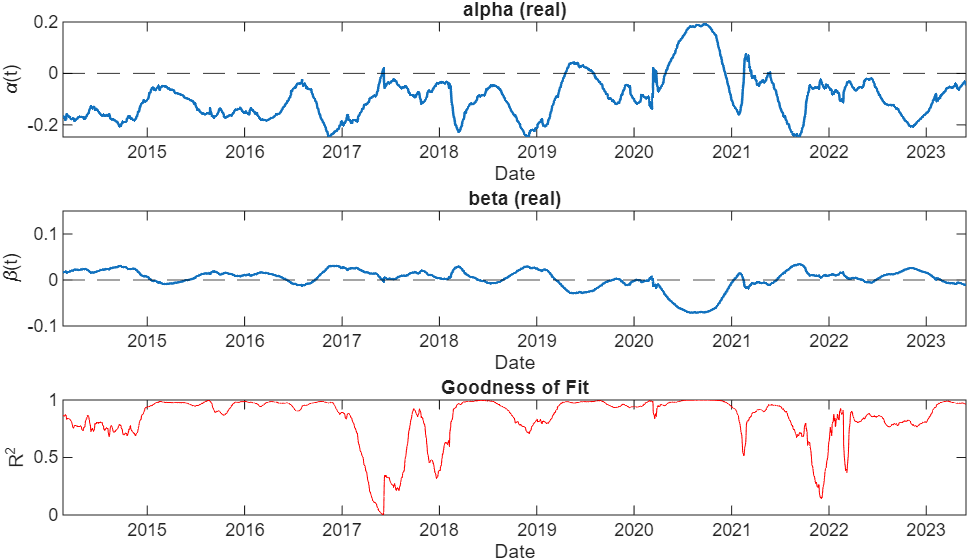}
\hfill
\includegraphics[
  width=0.48\textwidth,
  height=13cm,
  keepaspectratio
]{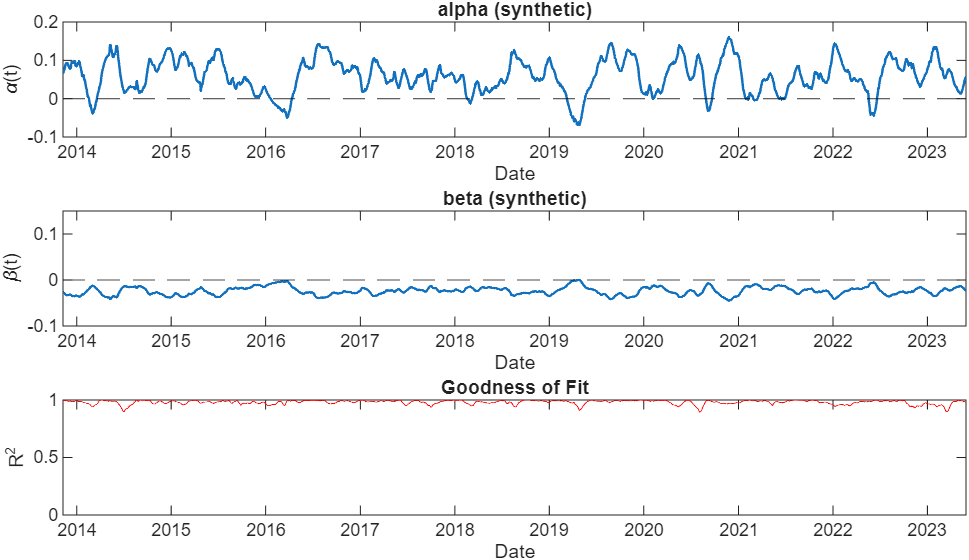}
\caption{First- and second-order coefficients ($\alpha(t)$, $\beta(t)$) and fit
quality ($R^2$) for real (left) and synthetic (right) data.}
\end{figure}

Real data display pronounced and time-varying curvature, whereas synthetic data
exhibit $\beta(t)\approx 0$, indicating an approximately linear $\log L$--$\log s$
relation.

The generalized effective exponent,
\begin{equation}
\delta_{\mathrm{eff}}(s,t) =
\frac{d\log L(s,t)}{d\log s}
= \alpha(t) + 2\beta(t)\log s,
\label{eq:effective_exponent}
\end{equation}
highlights scale-dependent shifts in real systems that are not captured by a
single, scale-independent exponent.

\section{Detecting multiscale network instabilities}

We adopt a simple deviation–screening approach to detect abrupt distortions in 
the multiscale network structure.  
Let $E_t$ denote the Elastic DCCR, defined using the following
two scales,

$s_{\mathrm{long}}=74$ (approximately 3.5 months) and $s_{\mathrm{short}}=11$ 
(approximately 2 weeks).  
We compute the $z$--score
\begin{equation}
Z_t = \frac{E_t - \mu_t}{\sigma_t},
\end{equation}
where $\mu_t$ and $\sigma_t$ are the 21-day rolling mean and standard deviation 
of $E_t$.  
Observations with $|Z_t|>2.5$ are classified as \emph{anomalies}.  
The threshold is used purely to flag exceptional local deviations without 
invoking any distributional assumptions.

\begin{figure}[H]
\centering
\includegraphics[width=0.85\textwidth]{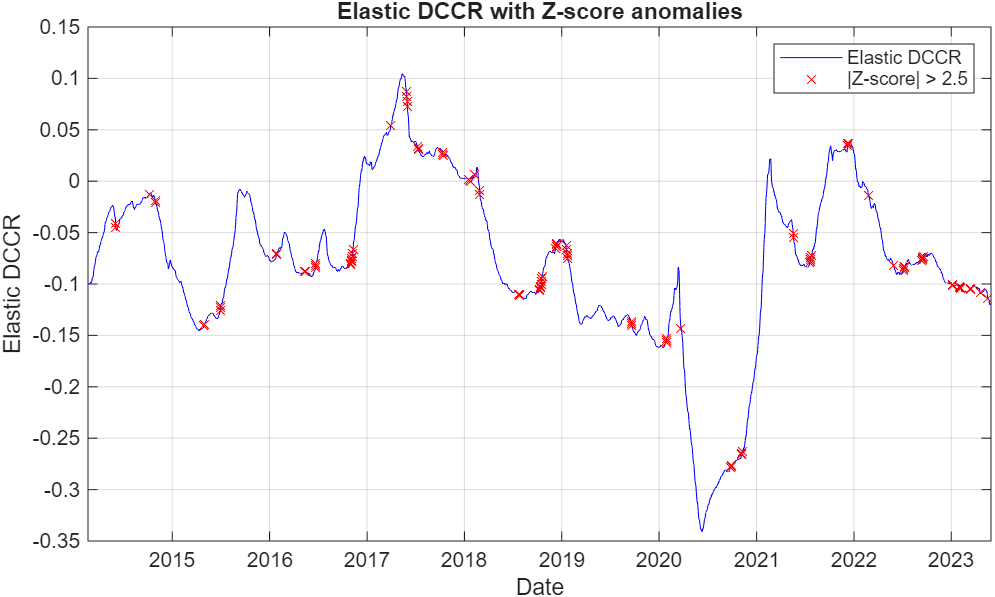}
\caption{$z$--score screening of the Elastic DCCR (ratio 74/11); anomalies 
defined by $|Z_t|>2.5$.}
\end{figure}

Points flagged in this manner correspond to short-lived but sharp transitions in 
the correlation-network topology.  
This complements volatility-based indicators by capturing structural deviations 
that are abrupt but not necessarily persistent.

\noindent\textbf{Event matching.}  
To assess the economic relevance of the detected anomalies, we examine clusters
of consecutive trading days with $|Z_t|>2.5$ (bridging weekends) and inspect
headline news within a $\pm 1$–day window.
 
Sources include international newswires (e.g., Reuters, Yahoo Finance) and 
policy calendars from major central banks.  
When several developments coincide with a given cluster, we prioritise 
internationally relevant events consistent with the global scope of the network.

This retrospective correspondence is judgement-based and hence subject to 
look-ahead bias.  
Future work could improve robustness by:  
(i) restricting the search to a pre-defined calendar of candidate events, and  
(ii) benchmarking overlaps against randomly shifted event dates.

\paragraph{Event correspondence}  

As shown in Table~\ref{tab:events}, the largest Elastic DCCR anomalies align with major global shocks.

\begin{table}[t]
\centering
\caption{Selected Elastic DCCR anomaly clusters and corresponding events.}
\label{tab:events}
\small
\setlength{\tabcolsep}{6pt}
\renewcommand{\arraystretch}{1.15}
\begin{tabular}{@{} l c l @{}}
\toprule
\textbf{Date range} & \textbf{Max $Z$} & \textbf{Event type} \\
\midrule
2015-06-25 -- 2015-06-26 & 2.62 & Greek capital controls \\
2016-06-20 -- 2016-06-23 & 2.84 & Brexit referendum \\
2017-04-03 -- 2017-04-04 & 2.75 & Article~50 negotiations \\
2020-01-27 -- 2020-01-28 & 2.58 & Wuhan lockdown \\
2020-11-09 -- 2020-11-10 & 2.59 & Pfizer--BioNTech vaccine \\
\bottomrule
\end{tabular}
\end{table}

The majority of anomaly clusters align with major macro-financial or geopolitical
events.
  
Examples include: Greece’s capital-control announcement (30 June 2015), the U.K. 
Brexit vote (24 June 2016), the Wuhan lockdown at the onset of COVID-19 
(27 January 2020), Pfizer–BioNTech’s vaccine-efficacy announcement 
(9 November 2020), the Federal Reserve’s “unlimited” QE (23 March 2020), and 
Gazprom’s suspension of \textit{Nord Stream~1} flows (1 September 2022).  
Only a few clusters lack an identifiable contemporaneous driver, indicating that 
the Elastic DCCR predominantly captures economically meaningful structural 
shifts rather than statistical artefacts.

\paragraph{Sign information}  
The sign of the deviation provides additional interpretation.  
\textbf{Positive} spikes ($Z_t>0$) typically coincide with periods of stress and 
risk-off compression in cross-market dependencies.  
\textbf{Negative} spikes ($Z_t<0$) tend to follow strong stabilisation signals 
(e.g., quantitative-easing announcements or vaccine breakthroughs) and reflect a 
rapid re-differentiation of correlations across scales.  
Although exceptions occur (e.g., the \textit{Nord Stream~1} shutdown yields a 
negative spike), the overall pattern supports interpreting $Z_t$ as a proxy for 
the balance between correlation compression and relaxation.

These episodes correspond to well-known global shocks associated with abrupt reconfigurations of cross-market dependencies.

\noindent\textbf{Interpretation.}  
Although individual return series exhibit comparable long-range memory
properties, the Elastic DCCR reveals scale-dependent reconfiguration of the
correlation network.
  
These deformations correspond to system-level adjustments triggered by
macroeconomic shocks, suggesting that the observed instabilities arise from
evolving interdependencies rather than marginal volatility dynamics.

The analysis of $L(s,t)$ across scales yields two main insights:
\begin{enumerate}
\item \textbf{Local scaling.}  
Log--log plots of $L(s,t)$ versus $s$ are approximately linear over limited scale
ranges at each time point.
This motivates the use of a time-varying scaling exponent $\alpha(t)$ as a
diagnostic of local structural complexity.
\item \textbf{Deviation from perfect scaling.}  
A curvature-based diagnostic indicates that the mean curvature of the log--log
profiles is nonzero.
Thus, although the relationship is locally linear, the system does not follow an 
exact power law globally.  
The persistent curvature is consistent with multiscale, and potentially
multifractal, organisation of the correlation network.
\end{enumerate}

\section{Comparison with the DCCA--GARCH approach of Diebold et~al.}

We compare our multiscale network methodology with the connectedness framework 
of Diebold et~al.~\cite{Diebold2014, Diebold2023}.  
Their approach measures directional and system-wide connectedness using 
generalized forecast-error variance decompositions (GVDs) from vector 
autoregressions (VARs).  
In this setting, connectedness reflects the extent to which shocks to one 
variable contribute to the forecast-error variance of another, thereby 
quantifying the propagation channels of system-wide risk.

Let $d^H_{ij}$ denote the $H$-step-ahead GVD component that attributes the 
forecast-error variance of variable $i$ to shocks in variable $j$ ($i\neq j$).  
The $N\times N$ connectedness table is then constructed from these 
non-own variance decompositions, with row and column sums representing 
directional “from”, “to”, and “net’’ connectedness.  
The framework uses the generalized variance-decomposition approach of 
Koop et~al.~\cite{koop1996impulse} and Pesaran and Shin~\cite{pesaran1998generalized}, 
which avoids the ordering dependence inherent in Cholesky-based orthogonalization 
and preserves the empirical correlation structure of shocks.

\begin{figure}[H]
\centering
\includegraphics[width=0.5\textwidth]{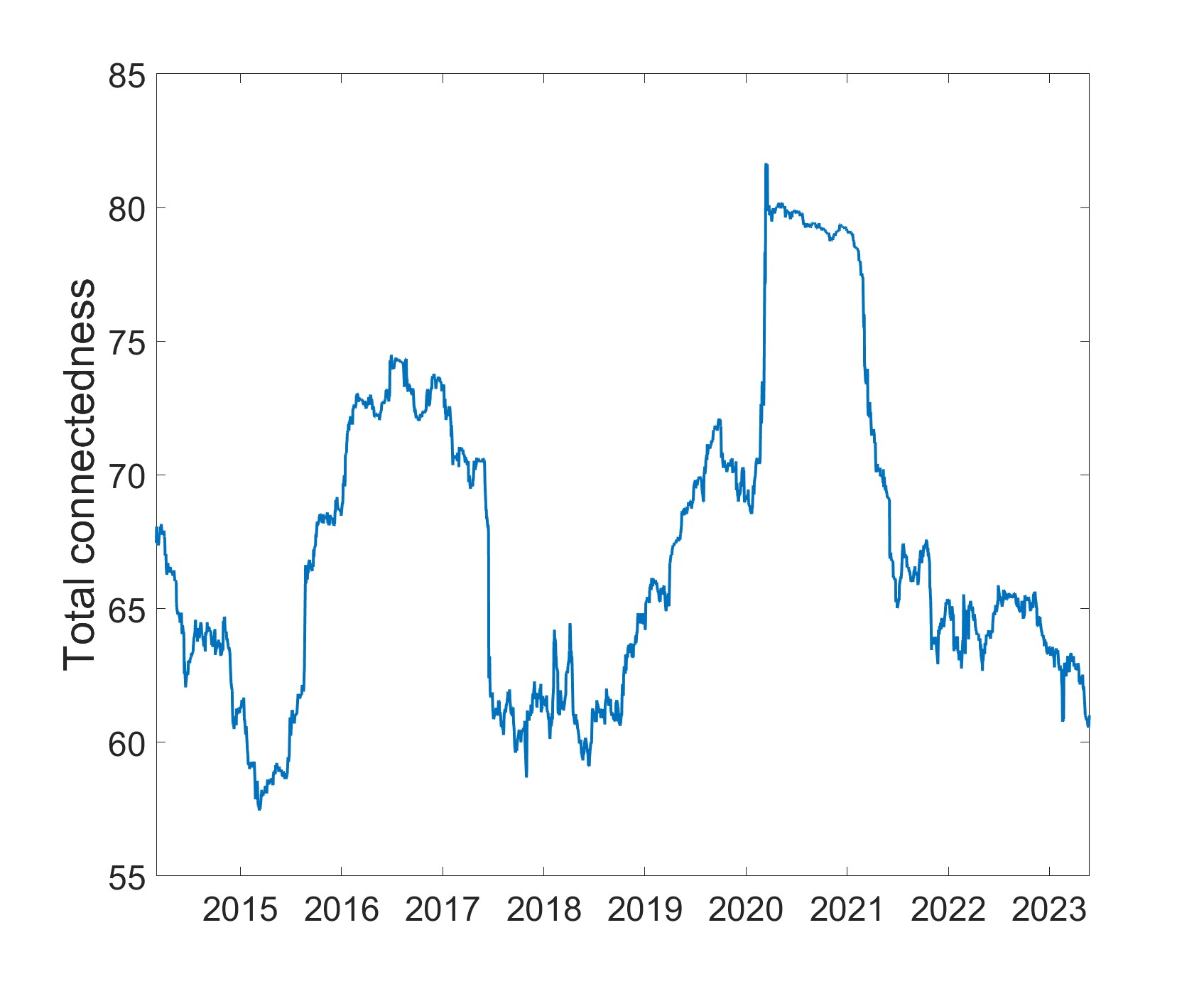}
\caption{Time-varying total connectedness in the sense of Diebold et~al., using 
a rolling window of $w=250$ days.}
\label{fig:Dieboldconn}
\end{figure}

Following Refs.~\cite{Diebold2014, Diebold2023}, we estimate a rolling VAR with 
window width $w=250$ trading days to obtain time-varying total connectedness.  
Figure~\ref{fig:Dieboldconn} shows the resulting series using the same dataset 
as in previous sections.  A full directional connectedness table is provided in 
Appendix~\ref{app:b}.

The Diebold et~al.\ total connectedness measure exhibits a strong empirical
correlation with the dominant eigenvalue of the DCCA-based adjacency matrix
(Pearson correlation $0.81$ for $w=250$), with similarly strong correlations
observed for other window lengths.

We compute “to”, “from”, and “net’’ directional components for completeness, and 
the code replicating all results is provided in the project repository.

\vspace{2mm}
\noindent\textbf{Conceptual differences.}  
Our approach differs from the Diebold et~al.\ framework in two key respects:

\begin{enumerate}
    \item \textbf{Cross-correlation metric.}  
    We transform scale-dependent DCCA coefficients into a geometric distance, 
    enabling the construction and monitoring of minimum spanning trees across 
    scales.  
    This provides a sparse, interpretable representation of evolving 
    cross-market structure.

   \item \textbf{Relaxation of covariance-stationarity.}
The DCCA-based methodology does not require covariance stationarity or VAR
specifications, allowing it to capture long-memory effects, slowly varying
dependencies, and multiscale structure that may be difficult to accommodate
within variance-decomposition frameworks.

\end{enumerate}

Thus, while the Diebold et~al.\ measure offers a VAR-based description of shock 
propagation, our framework provides a multiscale, topology-oriented view of 
connectedness that remains applicable in the presence of long-range dependence and structural
nonstationarity.

\section{Conclusions}

We introduce a multiscale framework for detecting structural transitions in
complex systems by analysing the evolution of cross-correlation networks.
The approach combines Detrended Cross-Correlation Analysis (DCCA) with
Minimum Spanning Tree (MST) filtering to obtain a sparse, multiscale
representation of interdependencies.
Its core component, the Elastic Detrended Cross-Correlation Ratio
(Elastic~DCCR), provides a dimensionless measure of how network topology
responds to changes in observational scale, thereby capturing local departures
from scale-invariant behaviour.

Applied to global equity indices, the method identifies clear episodes of
topological reorganisation coinciding with major financial disruptions.
These shifts manifest as short-horizon correlation compression and
long-horizon structural fragmentation, suggesting changes in the underlying
coordination of market components.
The Elastic~DCCR effectively isolates these transitions, revealing multiscale
adjustments that are not visible in single-scale or static measures of
dependence.

Because the framework relies solely on scale-dependent correlations and does not
require covariance stationarity, it is applicable to a broad class of systems
characterised by long-range dependence or nonstationary interactions.
Potential applications include climate networks, neuronal or biological systems,
and engineered infrastructures.
In this sense, the Elastic~DCCR provides a general and interpretable tool for
monitoring multiscale sensitivity to structural change and identifying emerging
instability regimes in complex interconnected systems.

\subsection*{Research perspectives and extensions}
\label{Research perspectives}

\subsubsection*{Scale-invariant cross-correlation under heterogeneous long-range dependence}

A natural extension of this work concerns the behaviour of DCCA-based correlation
measures when the underlying time series exhibit heterogeneous scaling
properties.
In many systems, the Hurst exponents $H_x$ and $H_y$ of two series differ
substantially, reflecting asymmetric persistence, multifractality, or
heterogeneous long-range memory.
Under such conditions, the standard DCCA normalization may no longer guarantee
boundedness or approximate scale invariance of $\rho_{\mathrm{DCCA}}(s)$,
particularly when the associated fluctuation functions grow at different rates.

One possible direction is a scaling-adjusted normalization that weights the DFA
fluctuation functions according to their relative scaling exponents,
\begin{equation}
\tilde{\rho}_{\mathrm{DCCA}}(s)
   = \frac{F^2_{\mathrm{DCCA}}(s)}
           {F_{\mathrm{DFA},x}(s)^\alpha
            F_{\mathrm{DFA},y}(s)^{1-\alpha}},
\end{equation}
with
\begin{equation}
\alpha = \frac{H_y}{H_x + H_y}.
\end{equation}

Assuming the scaling relations
\begin{equation}
F_{\mathrm{DFA},x}(s) \sim s^{H_x},\qquad
F_{\mathrm{DFA},y}(s) \sim s^{H_y},\qquad
F^2_{\mathrm{DCCA}}(s) \sim s^{2H_{xy}},
\end{equation}
the generalized coefficient behaves as
\begin{equation}
\tilde{\rho}_{\mathrm{DCCA}}(s)
   \sim s^{\,2H_{xy} - \left(\alpha H_x + (1-\alpha)H_y\right)}.
\end{equation}
Choosing $\alpha = H_y/(H_x+H_y)$ yields an exponent that coincides with $H_{xy}$
under idealized conditions, thereby attenuating spurious scale dependence.

While such an adjustment does not ensure strict boundedness in $[-1,1]$, it may
improve robustness when analysing systems with heterogeneous long-range
dependence.
For distance-based network applications, the resulting coefficient could be
further transformed or clipped to maintain metric compatibility.

\vspace{2mm}
\subsubsection*{Eigenvalue-based stability and multiscale network sensitivity}

The eigenvalue spectrum of a scale-dependent adjacency matrix provides an
additional diagnostic of structural stability in evolving networks.
In many dynamical systems, the dominant eigenvalue governs the sensitivity of
collective modes to perturbations and the onset of instability
\cite{sarkar2011spectral}.
A large leading eigenvalue indicates strong effective coupling, favouring rapid
propagation of shocks and reduced resilience.

In the present context, the dominant eigenvalue of the DCCA-based adjacency
matrix, or of its MST-filtered counterpart, offers a complementary measure of
system-wide coherence.
A systematic comparison between the Elastic~DCCR and the temporal evolution of
the leading eigenvalue may reveal whether breakdowns of approximate
scale-invariance coincide with increased spectral concentration.
Such an analysis could clarify how multiscale deformation of network topology is
linked to the emergence of collective instability modes.

Future work may therefore explore joint diagnostics combining:
(i) scale-dependent topological sensitivity (Elastic~DCCR), and
(ii) spectral stability indicators derived from the adjacency matrix.
This pairing has the potential to identify early-warning signatures of
instability in a wide class of multiscale networks.
\subsubsection*{Connection between scaling behaviour and cointegration}

A further research direction concerns the potential connection between the
scaling exponent $\alpha(t)$—and hence the Elastic~DCCR—and the concept of
cointegration.
If two price series are cointegrated, their log levels share a common stochastic
trend, while deviations from equilibrium remain stationary.
Within the DCCA--MST framework, such long-run co-movement is expected to manifest
as persistently high $\rho_{\mathrm{DCCA}}(s)$ at large scales~$s$, leading to
small DCCA distances
\[
d_{ij}(s)=\sqrt{2\!\left(1-\rho_{ij,\mathrm{DCCA}}(s)\right)}.
\]

Under an empirical scaling relation of the form $L(s,t)\propto s^{\alpha(t)}$,
this behaviour is consistent with values of $\alpha(t)$ that are close to zero
or negative, indicating that distances do not increase with scale.
Conversely, positive values of $\alpha(t)$ reflect dominance of short-term
co-movement and are consistent with the absence, or breakdown, of stable
cointegrating relationships.

From this perspective, $\alpha(t)$ and the Elastic~DCCR may be viewed as
nonparametric, scale-dependent diagnostics of long-run co-movement.
Persistent near-zero or negative values would be associated with stable
long-run linkages, whereas positive or highly volatile values may signal
fragmentation of equilibrium relationships, as often observed during crisis
periods.
Exploring this connection could provide a multiscale interpretation of
cointegration and help bridge traditional error-correction models with
geometric network representations.

\section*{Appendix A: Statistics of interest}
\label{app:a}
\setcounter{table}{0}
\renewcommand{\thetable}{A\arabic{table}}

Table~\ref{tab:statistics} reports descriptive statistics for the observed equity-index return series, including annualised mean and volatility, skewness, kurtosis, autocorrelation at various lags, and autocorrelation of squared returns.

\begin{table}[htbp]
\centering
\resizebox{\textwidth}{!}{
\begin{tabular}{lccccccccc}
\toprule
Stat\textbackslash Tickers & GSPC & GSPTSE & FCHI & GDAXI & FTSEMIB.MI & N225 & FTSE & HSI & IMOEX.ME \\
\midrule
Ann.\ mean & 0.10 & 0.04 & 0.06 & 0.07 & 0.05 & 0.09 & 0.02 & $-$0.02 & 0.06 \\
Ann.\ vol & 0.18 & 0.15 & 0.19 & 0.19 & 0.23 & 0.20 & 0.16 & 0.20 & 0.22 \\
Skewness & $-$0.82 & $-$1.66 & $-$0.81 & $-$0.58 & $-$1.41 & $-$0.19 & $-$0.88 & 0.04 & $-$2.25 \\
Kurtosis & 19.24 & 45.30 & 13.49 & 12.85 & 19.01 & 7.77 & 15.94 & 6.90 & 53.33 \\
AC 1d & $-$0.14*** & $-$0.09*** & $-$0.00 & $-$0.01 & $-$0.06*** & $-$0.02 & $-$0.00 & 0.01 & 0.09*** \\
AC 5d & 0.05*** & 0.04*** & 0.00 & 0.01 & 0.01*** & $-$0.01*** & 0.02 & $-$0.03 & $-$0.02*** \\
AC 10d & $-$0.06*** & $-$0.03*** & $-$0.03*** & $-$0.04*** & $-$0.01*** & $-$0.02*** & $-$0.00*** & $-$0.00 & $-$0.04*** \\
AC 20d & $-$0.02*** & 0.00*** & 0.02** & 0.03*** & $-$0.00** & $-$0.01*** & 0.04*** & 0.05* & $-$0.02*** \\
AC sq.\ ret 1d & 0.48*** & 0.43*** & 0.11*** & 0.06*** & 0.15*** & 0.21*** & 0.17*** & 0.28*** & 0.68*** \\
AC sq.\ ret 5d & 0.32*** & 0.28*** & 0.12*** & 0.08*** & 0.06*** & 0.08*** & 0.11*** & 0.13*** & 0.02*** \\
AC sq.\ ret 10d & 0.24*** & 0.14*** & 0.11*** & 0.09*** & 0.05*** & 0.14*** & 0.14*** & 0.09*** & 0.01*** \\
AC sq.\ ret 20d & 0.11*** & 0.06*** & 0.05*** & 0.07*** & 0.01*** & 0.03*** & 0.09*** & 0.06*** & 0.02*** \\
\bottomrule
\multicolumn{10}{l}{\small Notes: *, **, and *** denote significance at the 5\%, 1\%, and 0.1\% levels, respectively.}
\end{tabular}
}
\caption{\label{tab:statistics}Summary statistics for the equity-index return series.}
\end{table}

Tables~\ref{tab:1month} and~\ref{tab:4months} present full-sample DCCA coefficients at the 1- and 4-month time scales.

\begin{table*}[ht]
\centering
\begin{tabular}{ccccccccc}
\hline
 & \multicolumn{8}{c}{\textbf{DCCA coefficients ($s=1$ month)}}\\ \cline{2-9}
 & US & CA & FR & DE & IT & JP & UK & CN \\
\hline
CA & 0.8280 \\
FR & 0.7336 & 0.7491 \\
DE & 0.7350 & 0.7269 & 0.9416 \\
IT & 0.6543 & 0.6786 & 0.8995 & 0.8697 \\
JP & 0.6050 & 0.5814 & 0.6624 & 0.6429 & 0.5874 \\
UK & 0.6960 & 0.7548 & 0.8667 & 0.8215 & 0.7727 & 0.6054 \\
CN & 0.4948 & 0.4886 & 0.5283 & 0.5085 & 0.4645 & 0.5446 & 0.5302 \\
RU & 0.3948 & 0.4495 & 0.4649 & 0.4677 & 0.4328 & 0.3520 & 0.4584 & 0.3497 \\
\hline
\end{tabular}
\caption{Full-sample DCCA coefficients at the 1-month time scale.}
\label{tab:1month}
\end{table*}

\begin{table*}[ht]
\centering
\begin{tabular}{ccccccccc}
\hline
 & \multicolumn{8}{c}{\textbf{DCCA coefficients ($s=4$ months)}}\\ \cline{2-9}
 & US & CA & FR & DE & IT & JP & UK & CN \\
\hline
CA & 0.8584 \\
FR & 0.7972 & 0.7986 \\
DE & 0.7913 & 0.7745 & 0.9321 \\
IT & 0.6950 & 0.7287 & 0.9128 & 0.8625 \\
JP & 0.7212 & 0.6405 & 0.7217 & 0.7567 & 0.6484 \\
UK & 0.7869 & 0.8280 & 0.8734 & 0.8429 & 0.7786 & 0.6742 \\
CN & 0.4972 & 0.4731 & 0.5104 & 0.4895 & 0.4717 & 0.4481 & 0.5716 \\
RU & 0.4788 & 0.4997 & 0.5387 & 0.5514 & 0.5290 & 0.4510 & 0.5257 & 0.3614 \\
\hline
\end{tabular}
\caption{Full-sample DCCA coefficients at the 4-month time scale.}
\label{tab:4months}
\end{table*}

\newpage

\section*{Appendix B: Diebold et~al.\ measure of connectedness}
\label{app:b}
\setcounter{table}{0}
\renewcommand{\thetable}{B\arabic{table}}

Table~\ref{tab:example} reports the Diebold--Yilmaz connectedness measures computed for the dataset.

\begin{table}[htbp]
\centering
\resizebox{\textwidth}{!}{
\begin{tabular}{l*{10}{c}}
\hline
 & GSPC & GSPTSE & FCHI & GDAXI & FTSE.MIB & N225 & FTSE & HSI & IMOEX.ME & From \\
\hline
GSPC & 27.59 & 18.18 & 10.89 & 10.56 & 8.83 & 6.93 & 9.93 & 4.28 & 2.76 & 72.40 \\
GSPTSE & 18.34 & 27.39 & 18.86 & 10.02 & 9.41 & 4.71 & 11.75 & 3.87 & 3.60 & 72.60 \\
FCHI & 8.30 & 8.74 & 20.97 & 18.33 & 16.40 & 4.67 & 15.29 & 3.68 & 3.59 & 79.02 \\
GDAXI & 8.46 & 8.42 & 18.92 & 21.83 & 15.99 & 4.82 & 14.45 & 3.48 & 3.58 & 78.16 \\
FTSE.MIB & 7.82 & 8.53 & 17.98 & 16.96 & 23.98 & 4.42 & 13.53 & 3.19 & 3.53 & 76.01 \\
N225 & 5.69 & 5.52 & 7.42 & 6.43 & 5.52 & 51.29 & 7.39 & 8.55 & 2.15 & 48.70 \\
FTSE & 8.37 & 10.28 & 16.76 & 15.35 & 13.58 & 4.18 & 23.28 & 4.11 & 4.04 & 76.71 \\
HSI & 3.99 & 5.25 & 7.60 & 6.80 & 5.62 & 6.37 & 8.43 & 52.54 & 3.35 & 47.45 \\
IMOEX.ME & 4.63 & 6.52 & 8.40 & 8.08 & 7.47 & 2.71 & 8.56 & 3.28 & 50.31 & 49.68 \\
To & 65.64 & 71.47 & 98.88 & 92.56 & 82.86 & 38.85 & 89.37 & 34.47 & 26.64 & 66.75 \\
Net & $-$6.76 & $-$1.13 & 19.85 & 14.39 & 6.85 & $-$9.84 & 12.65 & $-$12.97 & $-$23.03 & 0 \\
\hline
\end{tabular}
}
\caption{\label{tab:example}Diebold--Yilmaz connectedness table for the full sample.}
\end{table}

\bibliography{sample}

\section*{Author contributions}
All authors contributed equally to this work. M.\,D., G.\,K., L.\,L.\ and J.\,DLM.\ participated in the study design, data analysis, interpretation and manuscript preparation. All authors approved the final version.

\section*{Additional information}
\textbf{Competing interests:} The authors declare no competing interests.

\end{document}